\documentclass[conference]{IEEEtran}
\IEEEoverridecommandlockouts

\usepackage{cite}
\usepackage{amsmath,amssymb,amsfonts}
\usepackage{graphicx}
\usepackage{textcomp}
\usepackage{graphicx}
\usepackage{colortbl}
\usepackage{subfig}
\usepackage{algorithm}
\usepackage{multirow}
\usepackage{booktabs}
\usepackage{subcaption}
\usepackage{url}
\usepackage{booktabs}
\usepackage{algorithm}
\usepackage{algpseudocode}
\usepackage[normalem]{ulem}
\useunder{\uline}{\ul}{}
\algnewcommand\algorithmicforeach{\textbf{for each}}
\algdef{S}[FOR]{ForEach}[1]{\algorithmicforeach\ #1\ \algorithmicdo}
\usepackage[table,xcdraw]{xcolor}
\usepackage[table,xcdraw]{xcolor}
\usepackage{xcolor}
\def\BibTeX{{\rm B\kern-.05em{\sc i\kern-.025em b}\kern-.08em
    T\kern-.1667em\lower.7ex\hbox{E}\kern-.125emX}}

\begin{document}

\title{Impact of a Sharpness Based Loss Function for Removing Out-of-Focus Blur\\
}

\author{
\IEEEauthorblockN{Uditangshu Aurangabadkar$^\star$, Darren Ramsook$^\dagger$, Anil Kokaram$^\ddagger$}
Sigmedia Group, Dept. of Electronic and Electrical Engineering,
{Trinity College Dublin, Ireland}\\
$\star$aurangau@tcd.ie, $\dagger$ramsookd@tcd.ie, $\ddagger$anil.kokaram@tcd.ie \\
\textit{www.sigmedia.tv}
}

\maketitle

\begin{abstract}
Recent research has explored complex loss functions for deblurring. In this work, we explore the impact of a previously introduced loss function -- $Q$ which explicitly addresses sharpness and employ it to fine-tune State-of-the-Art (SOTA) deblurring models. Standard image quality metrics such as PSNR or SSIM do not distinguish sharpness from ringing. Therefore, we propose a novel full-reference image quality metric $\Omega$ that combines PSNR with $Q$. This metric is sensitive to ringing artifacts, but not to a slight increase in sharpness, thus making it a fair metric for comparing restorations from deblurring mechanisms. Our approach shows an increase of 15\% in sharpness ($Q$) and up to 10\% in $\Omega$ over the use of standard losses. 
\end{abstract}


\section{Introduction}
\label{sec:introduction}

\renewcommand{\thefootnote}{\fnsymbol{footnote}}
\footnotetext[4]{This work was supported in part by a YouTube / Google Faculty award.}

Image deblurring is the problem of recovering a sharp image $\hat{I}$ from a blurry image $G$ such that it resembles the original image $I$. The process of blurring an image can be defined as follows:
\begin{equation}
    G = I * H_K + \eta
\end{equation}
where $H_K$ is a blur kernel of size $K \times K$ and $\eta \in {\cal N}(0, \sigma^2)$ is the additive Gaussian noise. Depending on whether the details of the blur kernel are given or not, the process can be classified as \textit{non-blind} or \textit{blind}, respectively. The current work focuses on blind deblurring as it is more relevant for real-world restoration tasks.

Classical methods such as the Wiener filter or Richardson-Lucy deconvolution~\cite{richardson1972bayesian, lucy1974iterative} have long been the standard for restoring images. However, with the widespread adoption of Deep Neural Networks (DNNs) for restoration, primarily deblurring, several works~\cite{kupyn2018deblurgan, zamir2022restormer} have proposed complex architectures that produce State-of-the-Art (SOTA) results in terms of visual quality of the reproduced image.

In practice, the main challenge in deblurring or restoration is to produce a sharper image without going so far as to introduce ringing artifacts. Ringing or \textit{over sharpening} results when high frequency components are boosted too much. A good restoration result is therefore often designed by applying a post-process after the initial deblurring step, removing any artifacts produced in the process.  Several works~\cite{vsroubek2019iterative, lopez2023deep, dong2020deep}, have now shifted to using a hybrid approach combining classical or iterative deblurring methods with Machine Learning (ML) based post-processors  to produce sharper restorations, yet carefully removing any artifacts produced in the process. The method proposed by Chen et. al.~\cite{chen2024deep} utilizes the Richardson-Lucy deconvolution algorithm as an initial step to remove blur from an image, followed by a Deep Neural Network (DNN) to remove any artifacts introduced in the process.

None of the previous techniques explicitly address sharpness of a restored image while also assessing the onset of ringing. What is needed, therefore, is a new loss function that focuses on enhancing this quality. An example would be to use the loss proposed by Aurangabadkar et. al.~\cite{10743912} that is based on a no-reference metric $Q$ proposed by Zhu and Milanfar~\cite{zhu2010automatic}, first introduced to optimize iterative denoising methods such as BM3D~\cite{4271520} with the goal of producing sharper restored images. In this paper, we combine this sharpness specific loss with the corresponding losses in a number of SOTA models in order to show the positive effect on image quality. This has not been done before.

An unfortunate consequence of using a loss function like $Q$, is that it increases monotonically with sharpness even when that sharpness manifests as ringing. 
An image with excessive ringing will therefore exhibit high $Q$ yet show a low Structural Similarity (SSIM)~\cite{wang2004image} or PSNR with respect to the ground truth reference. Inspired by this observation, we present a novel full-reference image quality metric $\Omega$ that combines PSNR with $Q$, such that it is sensitive to ringing. This new metric is then used to fairly assess the performance of our SOTA models trained with a hybrid loss function. Our primary contributions are as follows.
\begin{itemize}
    \item An analysis of the incorporation of our sharpness based loss function $Q$~\cite{10743912} in training SOTA deblurring models.
    \item A novel full-reference image quality metric that combines PSNR and metric $Q$, hence properly accounting for ringing. 
\end{itemize}

\section{Background}
It is clear that both the architectures and loss functions have an impact on the performance of DNN based deblurring mechanisms. We therefore consider three examples of architectures that span SOTA performance -- ARKNet~\cite{10743912}, XY-Deblur~\cite{ji2022xydeblur} and EHNet~\cite{ho2024ehnet}. 

ARKNet proposed by Aurangabadkar et. al.~\cite{10743912} is a standard U-Net~\cite{ronneberger2015u} based architecture comprising of 4 encoder \textit{layers}, where each layer consists of 5 convolutional blocks. Each block, in turn, comprises of a single $3 \times 3$ convolution layer, followed by Batch Normalization~\cite{ioffe2015batch} and GELU~\cite{hendrycks2016gaussian} activation. The model contains a total of 4.8 million trainable parameters. 

XY-Deblur introduced by Ji et. al.~\cite{ji2022xydeblur} is a single encoder multiple decoder architecture initially intended for restoring images degraded by motion blur. The model leverages the fact that employing multiple decoders allows for decomposing features into directional components, namely horizontal and vertical.  The use of shared kernels amongst the decoders allows for improved deblurring performance. These caveats keep the total number of trainable parameters identical to a standard U-Net, viz. 4.2 million, while producing significantly sharper restorations.  

EHNet proposed by Ho et. al.~\cite{ho2024ehnet} is a transformer based architecture that combines Convolutional Neural Networks (CNNs) and transformers to create a hybrid deblurring mechanism. The CNNs allow for efficient local feature extraction, whereas the transformer decoder with dual-attention enable the model to capture spatial and channel-wise dependencies. The model consists of approximately 8.7 million trainable parameters. Detailed diagrams of all the architectures can be found in the supplementary materials.~\footnote{https://github.com/aurangau/EUSIPCO2025}

\subsection{Loss Functions}
One way to address sharpness would be to use a loss function that explicitly increases sharpness in an image. In our previous work~\cite{10743912}, we propose a loss function which is based on a no-reference metric $Q$ introduced by Zhu and Milanfar~\cite{zhu2010automatic}. 

For measuring $Q$, an image is first divided into non-overlapping patches of size $k \times k$. The singular values $s_1$ and $s_2$ ($s_1 > s_2$), which determine the patch properties are calculated based on the eigen-decomposition of the matrix $\mathbf{G^{T}G}$, where $\mathbf{G}$ is the gradient matrix. Anisotropic patches (patches with texture) are then selected using an adaptive thresholding mechanism based on eigenvalues of $\mathbf{G^{T}G}$. $Q$ is then calculated as follows.
\begin{equation}
    Q = s_1 \cdot \frac{s_1 - s_2}{s_1 + s_2}
\end{equation}
Hence an image with sharper edges shows a large difference between eigenvalues from patches, producing a higher $Q$ than a blurry image. 

Each SOTA model discussed above employs a different loss function associated with that work. We wish to incorporate our sharpness based loss $Q$ into these loss functions. Therefore, denoting the existing loss for a particular SOTA model $\mathcal{L}_{\varphi}(\cdot)$, we deploy a composite loss augmenting this existing loss with our sharpness loss as follows.
\begin{equation}
    \mathcal{L} = \mathcal{L}_{\varphi}(I, \tilde{I}) - \beta \cdot Q(\tilde{I})  
    \label{eq:composite_loss}
\end{equation}
where $I$ and $\tilde{I}$ denote the Ground-Truth (GT) and restored images respectively. Note that $Q(\cdot)$ is a no-reference metric that depends only on the output deblurred image.

$\mathcal{L_\varphi}$ for ARKNet and XY-Deblur was ${l_1}$, whereas EHNet employs a combination of $l_1$ and frequency loss $\mathcal{L}_{freq}$ as follows.
\begin{equation}
    \mathcal{L}_{freq} = ||\mathcal{F}(\tilde{I}) - \mathcal{F}(I)||
    \label{eq:freq_loss}
\end{equation}
where $\mathcal{F}(\cdot)$ is the Fourier transform. 

\begin{figure}
    \centering
    \subfloat[PSNR=-; $\Omega$=-; $Q$=0.08]{\includegraphics[width=0.16\textwidth]{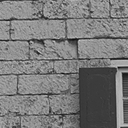}}\hfill
    \subfloat[PSNR=26.54; $\Omega$=1.59; $Q$=0.13]{\includegraphics[width=0.16\textwidth]{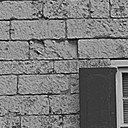}}\hfill
    \subfloat[PSNR=22.55; $\Omega$=5.85; $Q$=0.15]{\includegraphics[width=0.16\textwidth]{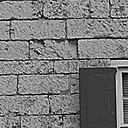}}\\
    \subfloat[PSNR=20.02; $\Omega$=11.09; $Q$=0.17]{\includegraphics[width=0.16\textwidth]{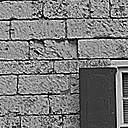}}\hfill
    \subfloat[PSNR=10.48; $\Omega$=10.98; $Q$=0.34]{\includegraphics[width=0.16\textwidth]{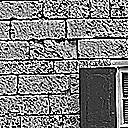}}\hfill
    \subfloat[PSNR=10.08; $\Omega$=10.50; $Q$=0.36]{\includegraphics[width=0.16\textwidth]{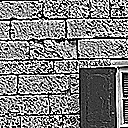}}
    \caption{\textit{Sensitivity of $\Omega$ to images with different sharpness amounts ($\gamma$)}. Top Row(L--R): Original Image. Image Sharpened with $\gamma$ = 0.8 and 1.3. Bottom Row(L--R): Image Sharpened with $\gamma$ = 2.5, 11.8 and 13.8. An increase in sharpness leads to an increase in $\Omega$, however with the introduction of ringing $\Omega$ decreases.}
    \label{fig:metric_omega_comparisons}
\end{figure}

To compare these systems it would be simple just to measure the loss function itself, but each method has a different effective loss function and so we need a fair metric. As stated previously, $Q$ increases with sharpness and also increases with ringing. However, PSNR or SSIM will reduce with ringing (therefore able to highlight that as a detrimental effect) but also reduces with sharpness. Figure ~\ref{fig:metric_omega_comparisons} shows this phenomenon by applying the Matlab function {\tt imsharpen} with parameter $\gamma$ to introduce sharpness into an image. We use here a $128 \times 128$ crop from an image in the Kodak dataset~\cite{franzen1999kodak}. We sharpen the original image with five different sharpness amounts $\gamma =  0.8, 1.3, 1.8, 2.5, 11.8, 13.8$. As sharpness ($\gamma$) increases, so does $Q$ but it continues to increase with ringing. PSNR on the other hand reduces with sharpness including when ringing artifacts appear. This effect is made more clear in Figure~\ref{fig:gamma_plot}.

What is needed therefore, is a measure that combines the strengths of standard picture quality metrics such as PSNR with $Q$, so that sharpness and ringing are treated correctly. We therefore propose a metric $\Omega$ which is a weighted combination of PSNR and $Q$. This is discussed next.

%

\section{Properly Accounting for Ringing}
As specified in the previous sections, an image with ringing will have a high $Q$, whilst having low SSIM or PSNR. We wish to devise a metric which increases with sharpness, but decreases with ringing artifacts in an image. We must note that PSNR behaves approximately correctly, but lacks the knowledge about sharpness, thus leading to a low value even for images that are sharp and have no spurious artifacts. $Q$, on the other hand, is a reliable indicator of sharpness when PSNR of a restored image is high, but becomes unreliable when PSNR is low. Therefore, we want to combine the two in a non-linear fashion given as follows.

\begin{equation}
    \Omega = (1 - \sigma(\alpha)) \cdot P' + \sigma(\alpha) \cdot \tilde{Q}
    \label{eq:metric_omega_patch}
\end{equation}

The intuition behind such a metric is based on the idea of giving more weight to PSNR for a patch that has high ringing and $Q$ when a patch has low ringing. The weighting function $\sigma(\cdot)$ is as follows.
\begin{equation}
    \sigma(\alpha) = \frac{1}{1 + e^{R(\alpha - \alpha_0)}}
    \label{eq:sigmoid_function}   
\end{equation}

\begin{figure}
    \centering
    \includegraphics[width=1.0\columnwidth]{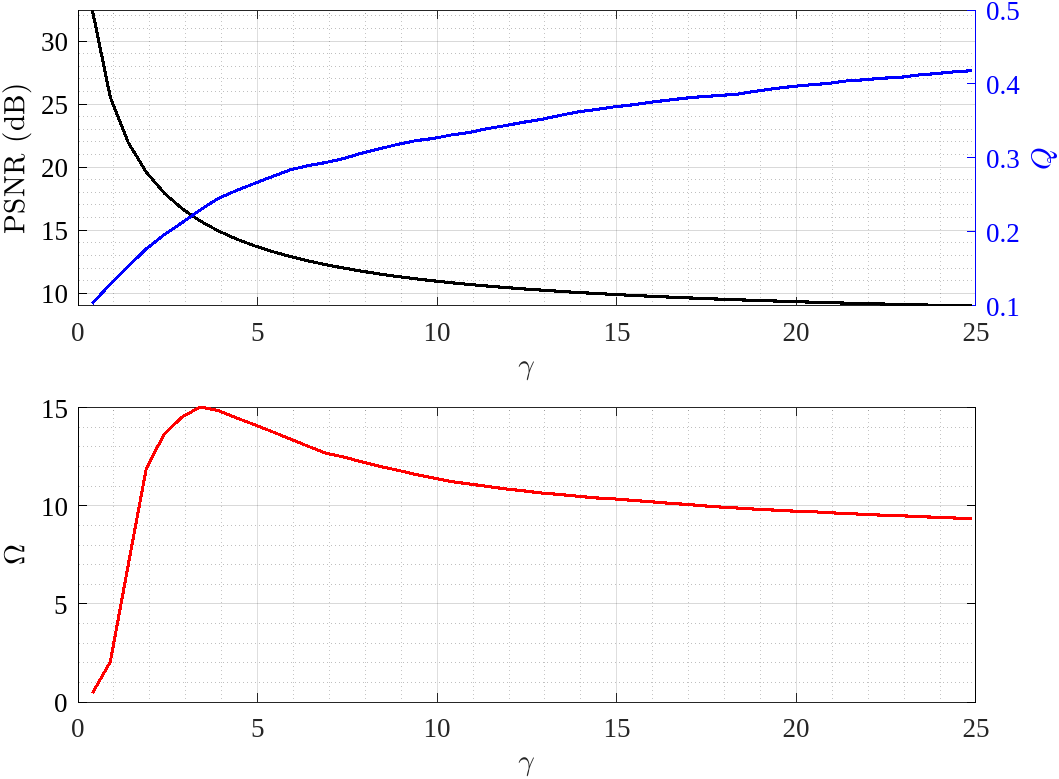}
    \caption{\textit{Behavior of $\Omega$, $Q$ and PSNR with increasing sharpness amounts ($\gamma$)}. Top: Comparison of $Q$ (blue) and PSNR (black) w.r.t $\gamma$. Bottom: Change in $\Omega$ w.r.t $\gamma$. As the sharpness increases, $Q$ increases, whereas PSNR decreases. However, $\Omega$ increases to a certain point and then begins to decrease as ringing is introduced.}
    \label{fig:gamma_plot}
\end{figure}

The parameters $R$ and $\alpha_0$ control the amount of \textit{acceptable} sharpness. Based on 13,052 images, we set $R$ = 5 and $\alpha_0$ = 1.2. Hence, an increase in sharpness leads to an increase in $\Omega$ and when ringing artifacts are produced, $\Omega$ decreases. Figure~\ref{fig:gamma_plot} shows a plot of the behavior of $\Omega$ PSNR and $Q$ with increasing amounts of $\gamma$. An increase in the sharpness amount ($\gamma$) initially leads to an increase in $\Omega$. However, as ringing artifacts are produced due to over-sharpening, $\Omega$ decreases. 

We divide the GT ($I$) and restored images ($\tilde{I}$) into $N$ non-overlapping patches of size $m \times m$. For each patch from both images, we measure $Q = Q(I)$ and $\tilde{Q} = Q(\tilde{I})$ respectively to determine the sharpness. To determine the amount of ringing, we measure a deviation ratio $\alpha$ given as follows:
\begin{equation}
    \alpha = \frac{|\tilde{Q} - Q|}{Q}
    \label{eq:alpha_calc}
\end{equation}

A higher $\alpha$ value corresponds to the presence of ringing in the restored patch. We then measure PSNR between the GT ($p$) and reference patch ($\tilde{p}$), clipping it at 50 dB, to avoid producing a value of $\infty$ when both patches are identical.
\begin{equation}
    P' = \min(PSNR(p, \tilde{p}), 50)
    \label{eq:psnr_calc}
\end{equation}
$\Omega$ for the entire image is the average per patch.

Figure~\ref{fig:metric_omega_comparisons} shows the behavior of $\Omega$ using a patch size $m=16$. As opposed to PSNR or $Q$, where the values are monotonically decreasing and increasing w.r.t the sharpness amounts, $\Omega$  increases with increasing sharpness for images with negligible ringing (Top Row, Center and Left). However, over-sharpening the image introduces ringing (Bottom Row), leading to a decrease in $\Omega$ values. 

\section{Experiments}
Our primary goal is to evaluate the loss $Q$ comprehensively. Therefore, we employ three different models and three different datasets. We train models with their respective losses $\cal L_\varphi$ for a fixed set of epochs until the best possible combination of weights are found and then fine-tune with our composite loss $\cal L$. The datasets used and the implementation specifics are mentioned next.

\subsection{Datasets}
\label{sec:datasets}
For training and testing the aforementioned methods, we use three datasets - RealDOF\cite{lee2021iterative}, DPDD\cite{abuolaim2020defocus} and ARK\cite{10743912}. The Real Depth of Field (\textbf{RealDOF}) set is a collection of 50 scenes, where each scene comprises of the in-focus image along with its out-of-focus counterpart. Each image is of the size $2320 \times 1536$ pixels. 

The Dual Pixel Defocus Dataset (\textbf{DPDD}) consists of 500 out-of-focus images along with their in-focus images. Each image is of the size $6720 \times 4480$ pixels. 

The dataset proposed by Aurangabadkar, Ramsook and Kokaram (\textbf{ARK}) comprises of three different out-of-focus images separated by the level of blur -- low, medium and high, captured by changing the focus ring of the camera by 2 units, w.r.t their in-focus Ground Truth (GT) counterparts. Each collection consists of 305 images of the size $5796 \times 3870$ pixels.

\subsection{Implementation Specifics}
For training the models, we use 143,104 crops from the DPDD dataset, 38,200 from the RealDOF dataset and 79,000 crops from the ARK dataset. For the purpose of inference, we use 4,996 crops from the RealDOF dataset, 7,898 crops from the DPDD dataset and 188 crops from the ARK dataset, all of which are of the size $128 \times 128$. Only the luma (Y) channel was used for deblurring.

The $\beta$ value in equation~\ref{eq:composite_loss} was empirically set to $0.1$ for training ARKNet, $0.01$ and $0.1$ for XY-Deblur and EHNet, respectively. Each model was trained with $\mathcal{L}_{\varphi}$ for 100 epochs and then fine-tuned using $\mathcal{L}$ for another 30 epochs. 

\section{Results}
For evaluating the quality of deblurred images, we use standard metrics such as PSNR, SSIM, alongside those that focus on sharpness -- $Q$, $J$~\cite{10647453}, $\Omega$, as well as, neural metrics such as LPIPS~\cite{zhang2018unreasonable}. $J$ is a sharpness metric normalized between [0, 1] that measures the sharpness difference of restored image $\hat{I}$ from the blurry and GT image. A restoration identical with the GT or blurry images will have a $J$ value of 1 or 0 respectively.

As can be seen from Table~\ref{tab:metrics_table}, incorporating $Q$ into the training leads to an increase of approximately 15\% in sharpness ($Q$) over using $\mathcal{L}_{\varphi}$ alone. We perform a paired t-test on the metric values and find the observed difference between means of metrics of images restored using $\mathcal{L}$ as opposed to $\mathcal{L_\varphi}$ are significant at the 5\% level ($p < 0.05$), the exception being SSIM, signifying its inability to properly detect sharpness in an image.

\begin{figure}
    \centering
    \includegraphics[width=1.0\columnwidth]{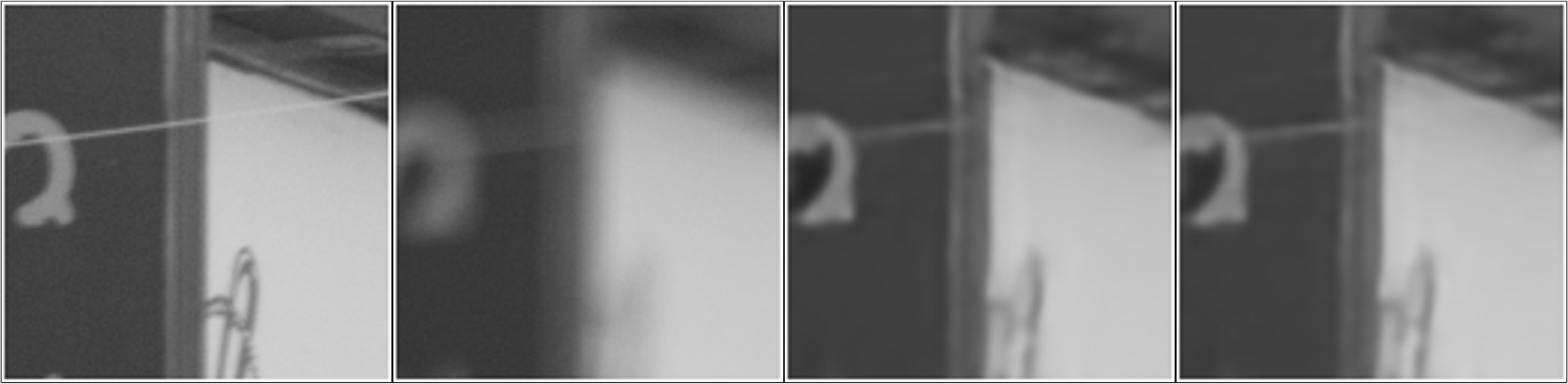}
    \caption{\textit{Ringing introduced by EHNet on an image from RealDOF dataset}. (L--R) GT image. Blurry image ($\Omega$=1.06). Image restored with $\mathcal{L_\varphi}$ ($\Omega$=1.75). Image restored with $\mathcal{L}$ ($\Omega$=1.72). Using our composite loss leads to the production of slightly over-sharpened edges, thus leading to a slight decrease in $\Omega$ value.}
    \label{fig:ringing_example}
\end{figure}

Our proposed metric $\Omega$ shows an increase of approximately 10\% for XY-Deblur and 3\% for ARKNet when we use ${\cal L}$ as opposed to ${\cal L}_\varphi$. averaged over three datasets For EHNet, however, we observe a slight reduction of 1\% when we use ${\cal L}$, as opposed to ${\cal L}_\varphi$, averaged over three datasets. This is possibly because of ringing. An example of this can be seen in Figure~\ref{fig:ringing_example}, where the model produces spurious artifacts around the edges. Therefore, subjective testing must be done to thoroughly understand the correlation between $\Omega$ and sharpness in an image. 

\begin{table}[]
\centering
\resizebox{\columnwidth}{!}{%
\begin{tabular}{@{}lccccccc@{}}
\toprule
\textbf{Method} &
  \textbf{Loss} &
  \textbf{PSNR $\uparrow$} &
  \textbf{SSIM $\uparrow$} &
  \textbf{$Q$ $\uparrow$} &
  \textbf{$J$ $\uparrow$} &
  \textbf{$\Omega$} &
  \textbf{LPIPS $\downarrow$} \\ \midrule
XY-Deblur~\cite{ji2022xydeblur} &                                         & 28.091 & 0.782 & { 1.229} & 0.642 & 3.093 & 0.388 \\
ARKNet~\cite{10743912}    &                                         & 28.050 & 0.778 & { 1.132} & 0.598 & 3.016 & 0.388 \\
EHNet~\cite{ho2024ehnet}     & \multirow{-3}{*}{$\mathcal{L}_\varphi$} & 27.992 & 0.773 & { 1.179} & 0.605 & 3.285 & 0.400 \\ \midrule
XY-Deblur &                                         & 27.930 & 0.780 & {\color[HTML]{FE0000} 1.518} & 0.699 & 3.409 & 0.394 \\
ARKNet    &                                         & 27.860 & 0.777 & {\color[HTML]{FE0000} 1.283} & 0.647 & 3.116 & 0.393 \\
EHNet     & \multirow{-3}{*}{$\mathcal{L}$}         & 28.065 & 0.777 & {\color[HTML]{FE0000} 1.211} & 0.622 & 3.250 & 0.394 \\ \bottomrule
\end{tabular}%
}
\caption{\textit{Comparison of incorporating $Q$ as a loss for 3 deblurring models averaged over 3 different datasets}. We see an increase of about 15\% in sharpness ($Q$) for methods where $\mathcal{L}$ was used (highlighted in red), as compared to those where $\mathcal{L}_\varphi$ was used. In contrast, PSNR, SSIM and LPIPS remain roughly the same, implying that $Q$ as a loss has targeted sharpness effectively.}
\label{tab:metrics_table}
\end{table}

\begin{figure}
    \centering
    \includegraphics[width=1.0\columnwidth]{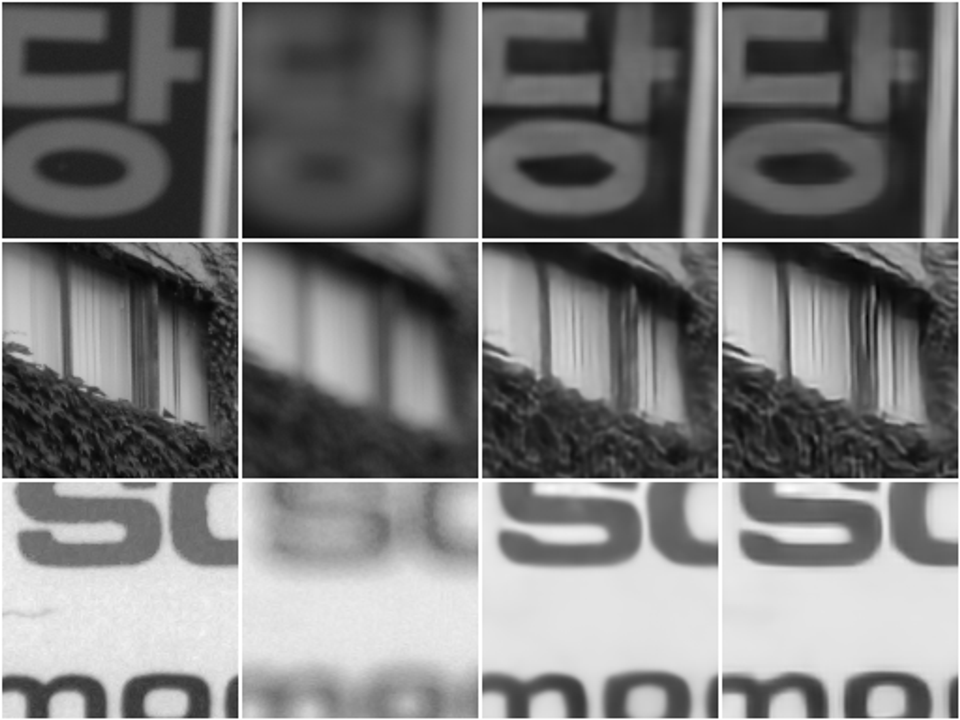}
    \caption{\textit{Comparison of restorations from XY-Deblur on 3 datasets}. (L--R) GT image. Blurry image ($\Omega$=3.07,1.47,1.04). Image restored with $\mathcal{L_\varphi}$ ($\Omega$=1.34,1.33,2.24). Image restored with $\mathcal{L}$ ($\Omega$=0.79,2.59,2.13). Models fine-tuned using our composite loss $\mathcal{L}$ produce sharper edges.}
    \label{fig:XYD_comparison}
\end{figure}

Figure~\ref{fig:XYD_comparison} provides a comparison of images from three different datasets  (rows 1 to 3) mentioned in section~\ref{sec:datasets} deblurred using XY-Deblur with and without $Q$ (columns 4 and 3, respectively). Using $Q$ during training produces noticeably sharper edges. 

\begin{figure}
    \centering
    \includegraphics[width=1.0\columnwidth]{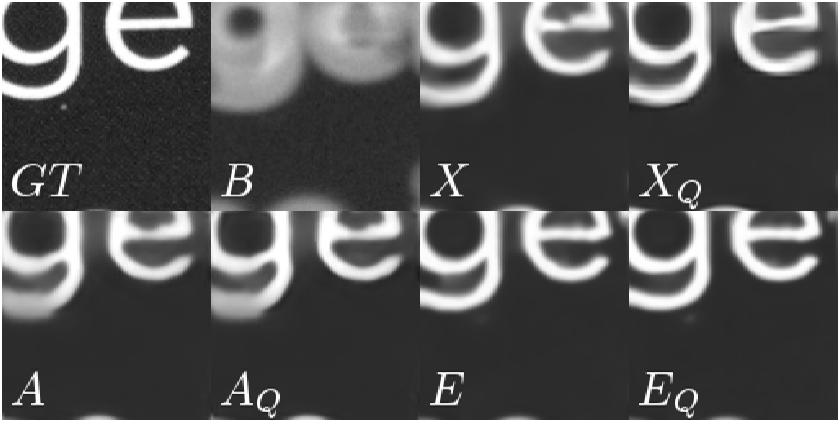}
    \caption{\textit{Visual comparison of different models without and with $Q$ on an image from the ARK dataset}. XY-Deblur, ARKNet and EHNet trained using $\mathcal{L}_\varphi$ alone are denoted by $X$, $A$ and $E$ respectively. The methods which were fine-tuned with $Q$ are denoted by $X_Q$, $A_Q$ and $E_Q$. All models produce finer textural information when $Q$ is incorporated into training.}
    \label{fig:model_comparison}
\end{figure}

Figure~\ref{fig:model_comparison} provides a comparison of the different methods trained with and without $Q$. Figures denoted by $GT$ and $B$ are the original and blurry images, respectively. All models trained using $\mathcal{L}$ produce much sharper restorations. Restorations generated by the complex transformer model EHNet augmented with our loss i.e. using ${\cal L}$ produce sharper images than other models.

\begin{figure}
    \centering
    \includegraphics[width=1.0\columnwidth]{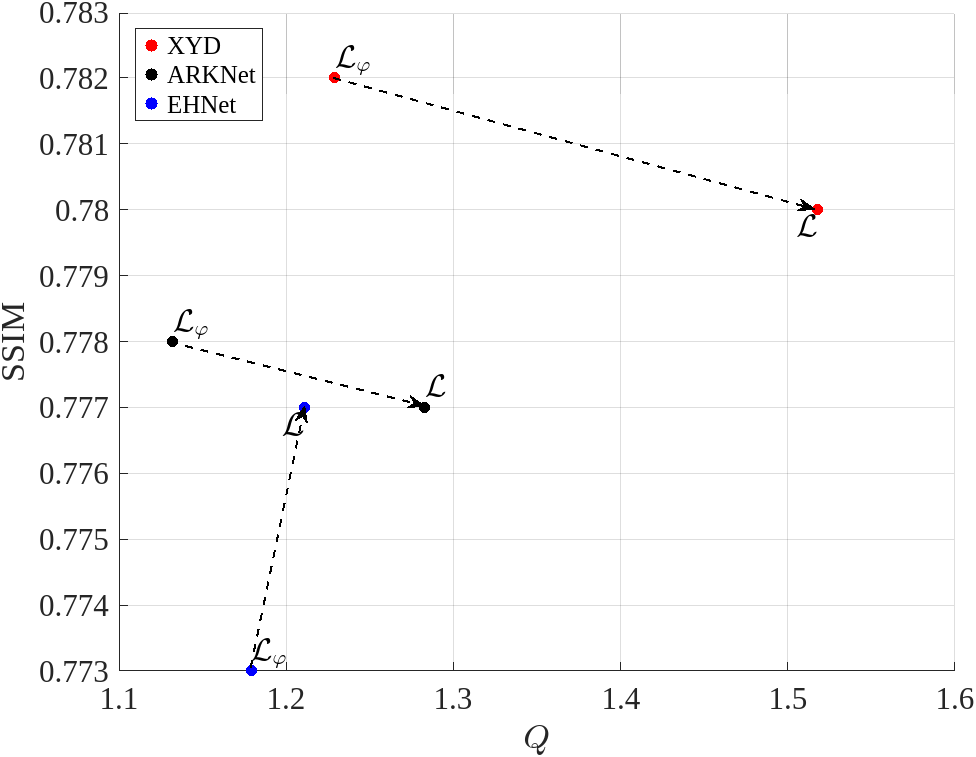}
    \caption{\textit{Trade-off between SSIM and $Q$}. Note that while the changes in $Q$ are statistically significant, those in SSIM are not.}
    \label{fig:tradeoff_comparison}
\end{figure}

Figure~\ref{fig:tradeoff_comparison} shows the effect of incorporating sharpness with respect to SSIM. Each point shows the average performance over the three datasets used. In each case the use of ${\cal L}$ causes an increase in sharpness (i.e. points displaced to the right on the plot) with negligible effect on SSIM. We must highlight that the changes in $Q$ are statistically significant, as opposed to SSIM. 

\section{Conclusion}
We have explored the impact of the loss function $Q$ in SOTA models to produce sharper restorations. We propose a novel full-reference metric $\Omega$ that is a weighted combination of PSNR with $Q$ and is sensitive to ringing artifacts.  Our approach shows an increase of approximately  15\% in sharpness ($Q$) and up to 10\% in $\Omega$ values of restored images, as compared to using losses that do not explicitly call out sharpness in images. We acknowledge that there is a scaling issue in $\Omega$. Essentially, $\Omega$ mixes decibels (dB) with a dimensionless measure ($Q$), thus implying that the range of this metric is content specific. While not an issue for optimization, it is an issue for interpretability of the metric. Our future work involves addressing this issue and using $\Omega$ as a loss in such networks and performing a large scale subjective study to investigate the correlation between $Q$, $J$ and $\Omega$ with Mean Opinion Scores. 

\bibliographystyle{IEEEbib}
\bibliography{refs}

\begin{thebibliography}{10}

\bibitem{richardson1972bayesian}
William~Hadley Richardson,
\newblock ``Bayesian-based iterative method of image restoration,''
\newblock {\em JoSA}, vol. 62, no. 1, pp. 55--59, 1972.

\bibitem{lucy1974iterative}
Leon~B Lucy,
\newblock ``An iterative technique for the rectification of observed distributions,''
\newblock {\em Astronomical Journal, Vol. 79, p. 745 (1974)}, vol. 79, pp. 745, 1974.

\bibitem{kupyn2018deblurgan}
Orest Kupyn, Volodymyr Budzan, Mykola Mykhailych, Dmytro Mishkin, and Ji{\v{r}}{\'\i} Matas,
\newblock ``Deblurgan: Blind motion deblurring using conditional adversarial networks,''
\newblock in {\em Proceedings of the IEEE conference on computer vision and pattern recognition}, 2018, pp. 8183--8192.

\bibitem{zamir2022restormer}
Syed~Waqas Zamir, Aditya Arora, Salman Khan, Munawar Hayat, Fahad~Shahbaz Khan, and Ming-Hsuan Yang,
\newblock ``Restormer: Efficient transformer for high-resolution image restoration,''
\newblock in {\em Proceedings of the IEEE/CVF conference on computer vision and pattern recognition}, 2022, pp. 5728--5739.

\bibitem{vsroubek2019iterative}
Filip {\v{S}}roubek, Tom{\'a}{\v{s}} Kerepeck{\`y}, and Jan Kamenick{\`y},
\newblock ``Iterative wiener filtering for deconvolution with ringing artifact suppression,''
\newblock in {\em 2019 27th European Signal Processing Conference (EUSIPCO)}. IEEE, 2019, pp. 1--5.

\bibitem{lopez2023deep}
Santiago L{\'o}pez-Tapia, Javier Mateos, Rafael Molina, and Aggelos~K Katsaggelos,
\newblock ``Deep robust image restoration using the moore-penrose blur inverse,''
\newblock in {\em 2023 IEEE International Conference on Image Processing (ICIP)}. IEEE, 2023, pp. 775--779.

\bibitem{dong2020deep}
Jiangxin Dong, Stefan Roth, and Bernt Schiele,
\newblock ``Deep wiener deconvolution: Wiener meets deep learning for image deblurring,''
\newblock {\em Advances in Neural Information Processing Systems}, vol. 33, pp. 1048--1059, 2020.

\bibitem{chen2024deep}
Liang Chen, Jiawei Zhang, Zhenhua Li, Yunxuan Wei, Faming Fang, Jimmy Ren, and Jinshan Pan,
\newblock ``Deep richardson--lucy deconvolution for low-light image deblurring,''
\newblock {\em International Journal of Computer Vision}, vol. 132, no. 2, pp. 428--445, 2024.

\bibitem{10743912}
Uditangshu Aurangabadkar, Darren Ramsook, and Anil Kokaram,
\newblock ``A sharpness based loss function for removing out-of-focus blur,''
\newblock in {\em 2024 IEEE 26th International Workshop on Multimedia Signal Processing (MMSP)}, 2024, pp. 1--6.

\bibitem{zhu2010automatic}
Xiang Zhu and Peyman Milanfar,
\newblock ``Automatic parameter selection for denoising algorithms using a no-reference measure of image content,''
\newblock {\em IEEE transactions on image processing}, vol. 19, no. 12, pp. 3116--3132, 2010.

\bibitem{4271520}
Kostadin Dabov, Alessandro Foi, Vladimir Katkovnik, and Karen Egiazarian,
\newblock ``Image denoising by sparse 3-d transform-domain collaborative filtering,''
\newblock {\em IEEE Transactions on Image Processing}, vol. 16, no. 8, pp. 2080--2095, 2007.

\bibitem{wang2004image}
Zhou Wang, Alan~C Bovik, Hamid~R Sheikh, and Eero~P Simoncelli,
\newblock ``Image quality assessment: from error visibility to structural similarity,''
\newblock {\em IEEE transactions on image processing}, vol. 13, no. 4, pp. 600--612, 2004.

\bibitem{ji2022xydeblur}
Seo-Won Ji, Jeongmin Lee, Seung-Wook Kim, Jun-Pyo Hong, Seung-Jin Baek, Seung-Won Jung, and Sung-Jea Ko,
\newblock ``Xydeblur: divide and conquer for single image deblurring,''
\newblock in {\em Proceedings of the IEEE/CVF conference on computer vision and pattern recognition}, 2022, pp. 17421--17430.

\bibitem{ho2024ehnet}
Quoc-Thien Ho, Minh-Thien Duong, Seongsoo Lee, and Min-Cheol Hong,
\newblock ``Ehnet: Efficient hybrid network with dual attention for image deblurring,''
\newblock {\em Sensors}, vol. 24, no. 20, pp. 6545, 2024.

\bibitem{ronneberger2015u}
Olaf Ronneberger, Philipp Fischer, and Thomas Brox,
\newblock ``U-net: Convolutional networks for biomedical image segmentation,''
\newblock in {\em Medical image computing and computer-assisted intervention--MICCAI 2015: 18th international conference, Munich, Germany, October 5-9, 2015, proceedings, part III 18}. Springer, 2015, pp. 234--241.

\bibitem{ioffe2015batch}
Sergey Ioffe,
\newblock ``Batch normalization: Accelerating deep network training by reducing internal covariate shift,''
\newblock {\em arXiv preprint arXiv:1502.03167}, 2015.

\bibitem{hendrycks2016gaussian}
Dan Hendrycks and Kevin Gimpel,
\newblock ``Gaussian error linear units (gelus),''
\newblock {\em arXiv preprint arXiv:1606.08415}, 2016.

\bibitem{franzen1999kodak}
Rich Franzen,
\newblock ``Kodak lossless true color image suite,''
\newblock {\em source: http://r0k. us/graphics/kodak}, 1999.

\bibitem{lee2021iterative}
Junyong Lee, Hyeongseok Son, Jaesung Rim, Sunghyun Cho, and Seungyong Lee,
\newblock ``Iterative filter adaptive network for single image defocus deblurring,''
\newblock in {\em Proceedings of the IEEE/CVF conference on computer vision and pattern recognition}, 2021, pp. 2034--2042.

\bibitem{abuolaim2020defocus}
Abdullah Abuolaim and Michael~S Brown,
\newblock ``Defocus deblurring using dual-pixel data,''
\newblock in {\em European Conference on Computer Vision}. Springer, 2020, pp. 111--126.

\bibitem{10647453}
Uditangshu Aurangabadkar and Anil Kokaram,
\newblock ``A dictionary based approach for removing out-of-focus blur,''
\newblock in {\em 2024 IEEE International Conference on Image Processing (ICIP)}, 2024, pp. 1494--1499.

\bibitem{zhang2018unreasonable}
Richard Zhang, Phillip Isola, Alexei~A Efros, Eli Shechtman, and Oliver Wang,
\newblock ``The unreasonable effectiveness of deep features as a perceptual metric,''
\newblock in {\em Proceedings of the IEEE conference on computer vision and pattern recognition}, 2018, pp. 586--595.

\end{thebibliography}

\end{document}